\documentclass[5p,times,final,a4paper,compress]{elsarticle}
\usepackage[utf8]{inputenc}
\usepackage{amssymb}
\usepackage{latexsym}
\usepackage{amsmath}

\usepackage[switch]{lineno}

\usepackage{booktabs}
\usepackage{xcolor,soul,ulem}
\usepackage[hidelinks]{hyperref}
\hypersetup{
    colorlinks,
    linkcolor={red!50!black},
    citecolor={blue!50!black},
    urlcolor={black}
}
\usepackage[labelsep=period]{caption}
\definecolor{revAcolor}{HTML}{000000}

\definecolor{revBcolor}{HTML}{000000}

\journal{Physical Review Letters}

\begin{document}
\begin{frontmatter}
\title{%
From Flow to Jamming: \\ Lattice Gas Automaton Simulations in Granular Materials
}
\author[gda1]{Mohamed Gaber\corref{cor1}}
\address[gda1]{Minerva University, 14 Mint Plaza, Suite 300, San Francisco, CA~94103, USA}
\ead{gaber@uni.minerva.edu}
\cortext[cor1]{Corresponding author}

\author[gda1]{Raquel H. Ribeiro}

\author[gda3,gda4]{Janek Kozicki}
\address[gda3]{Faculty of Applied Physics and Mathematics, Gdańsk University of Technology, 80-233 Gdańsk, Poland}
\address[gda4]{Advanced Materials Center, Gdańsk University of Technology, 80-233 Gdańsk, Poland}

\date{\today}


\begin{abstract}
\noindent
We introduce the first extension of a Lattice Gas Automaton (LGA) model to accurately replicate observed emergent phenomena in granular materials with a special focus on previously unexplored jamming transitions by incorporating gravitational effects, energy dissipation in particle collisions, and wall friction. We successfully reproduce flow rate evolution, density wave formation, and jamming transition observed in experiments. We also explore the critical density at which jamming becomes probable. This research advances our understanding of granular dynamics and offers insights into the jamming behavior of granular materials.

\end{abstract}
\end{frontmatter}

\section{\label{sec:level1} Introduction}

Granular materials, composed of solid particles like sand or grains, challenge conventional matter classifications~\cite{Shinbrot2000, Savage1989, Chung2010, BuiltOnSand2020}. Unlike regular substances, they mimic solid~\cite{Fagert1989, Herrmann1996}, liquid~\cite{Savage1989, BuiltOnSand2020, Kadanoff1989}, or gas states~\cite{Mesri2009} depending on the environment conditions~\cite{BuiltOnSand2020, Behringer1990, Fagert1989}, even forming a distinct matter phase (or a transitional bridge) between the primary phases~\cite{Chung2010}. Their unique attributes (such as insensitivity to temperature changes and nonlinear friction~\cite{Yang2012}) 
challenge predictive modeling ~\cite{Jaeger1989}, and yet offer the promise of innovative applications in the energy and pharmaceutics sectors~\cite{Chung2010}.

The boundary between solid- and fluid-like states in granular flow is marked by the \textit{jamming transition}, a well-studied shift from a flowing to a static state
~\cite{Jamming2001, Sadjadpour2008, Trappe2001, OHern2003, Teitel2007}, triggered by critical control parameters: density, temperature, and shear stress \cite{OHern2003}. This process resembles the glass transition in amorphous materials, 
from dynamic to ordered states~\cite{OHern2003}. 
Increased density naturally halts dynamics in granular materials, leading to the static, ``jammed'' state~\cite{Trappe2001, OHern2003, Teitel2007, PajicLijakovic2021, Jean2009, Giraldi2005}. Teitel et al. linked density to viscosity, showing viscosity rises with density until a critical point, triggering the solid jammed state~\cite{Teitel2007}. Shear stress and temperature further influence this state---when either falls below a critical threshold, the reduced mobility leads to the system becoming ``jammed.''

Density waves emerge as spatial density variations that reveal flow dynamics and mechanical traits~\cite{Fagert1989, Riethmuller1997}, and are therefore especially challenging to model. They are triggered by vibrations and shear forces imprinting intricate patterns and structures~\cite{Silbert2003, ElShamy2014}, and collectively contributing to particle motion.

This study leverages Lattice Gas Automaton (LGA) simulations 
to extend prior experimental work on density waves and jamming transition phenomena. 
Our novel contribution lies in the simulation of the jamming process by specifying appropriate model rules, addressing an unexplored gap in prior work 
and triggering potential practical applications \cite{Mukwiri2017}.

\section{\label{sec:level2} Model Description}
Modeling fluid flow 
dynamics requires solving the Navier--Stokes equations---a well-known computationally intensive task \cite{Judice2018, Matsushima1994, Hughes1998, PericNDE}.
In contrast, the Lattice Gas Automaton (LGA) model introduced by Frisch et al.~\cite{Frisch1986} offers a more efficient approach. Unlike traditional methods that require the definition of macroscopic variables and solving partial differential equations, LGA uses rules valid at microscopic scales to predict macroscopic behavior. This bottom-up approach simplifies continuous equations into discrete rules, rendering LGA an accurate model for fluid simulation~\cite{Chopard2009}. Notably, LGA effectively models granular materials, aligning closely with experimental observations \cite{ElShamy2014, Kozicki2005, Zhao2008, Fitt1992}. In this section, we present a modified LGA variant tailored for simulating density waves and jamming transitions in granular materials.

\subsection{\label{subsec:model-config} Model Configuration}

\begin{figure}[t]
  \includegraphics[width=\columnwidth]{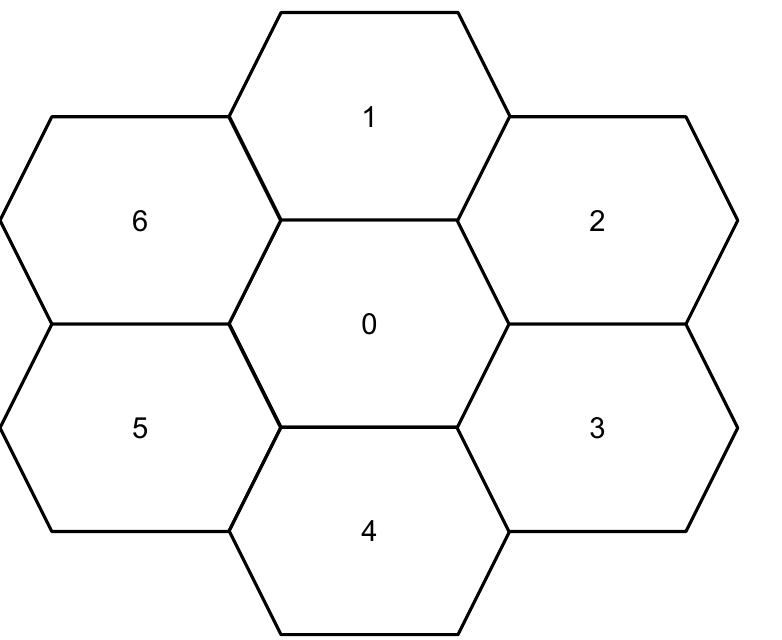}
  \caption{\label{fig:1} The cells are tagged with the particle's velocity. A value of $0$ indicates a particle at rest, while values $1-6$ represent particles in movement.}
\end{figure}

The LGA model comprises a $2D$ lattice with $L \times L$ hexagonal cells, each having six neighboring cells. Each cell can be occupied by one or more particles, empty, or act as a wall. The particles possess velocity vectors indicating their movement directions. The velocity vector consists of six binary elements, with only one element set to $1$ representing movement in the direction of that element (Fig.~\ref{fig:1}).

\subsection{\label{subsec:gravitational-effects} Adding Gravitational Effects}

We follow the approach by Kozicki et al.~\cite{Kozicki2005} and introduce a  parameter $g$, ranging from $0$ to $1$, which represents the probability of particles changing their velocity vector direction toward gravity at each iteration. For example, if the particle's direction is $0$ (Fig.~\ref{fig:1}), its new direction value will be $4$ with probability $g$. If the particle's direction were $2$, its new direction would be $3$ with probability $g$. This modification allows for parabolic behavior in individual particle velocities, consistent with the findings of Kozicki et al.~\cite{Kozicki2005}.

\subsection{\label{subsec:energy-dissipation} Adding Energy Dissipation}

\begin{figure}[t]
  \includegraphics[width=\columnwidth]{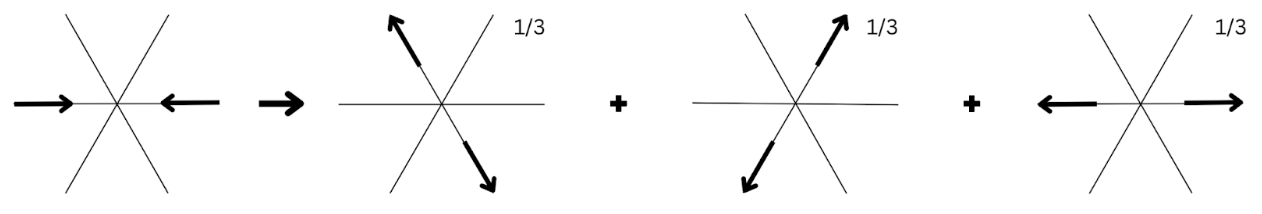}
  \caption{\label{fig:2} Head-on particle collisions result in one of three deflection outcomes, each with a 1/3 probability.}
\end{figure}

\begin{figure}[t]
  \includegraphics[width=\columnwidth]{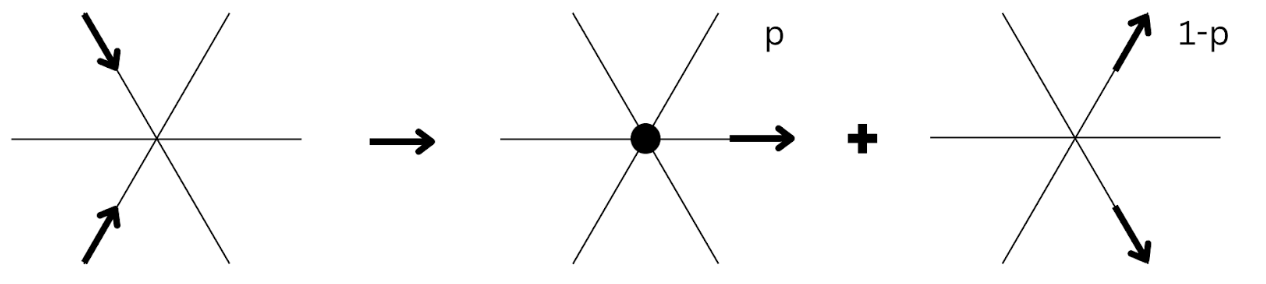}
  \caption{\label{fig:3} Collisions between particles at an angle yield different outcomes depending on the energy dissipation parameter, $p$. For $p = 0$, collision conservation laws govern energy and momentum. With $p > 0$, one of the particles loses velocity through energy dissipation. Note that given the hexagonal grid, the collision angles are equal to the rebound angles.}
\end{figure}

\begin{figure}[t]
  \includegraphics[width=\columnwidth]{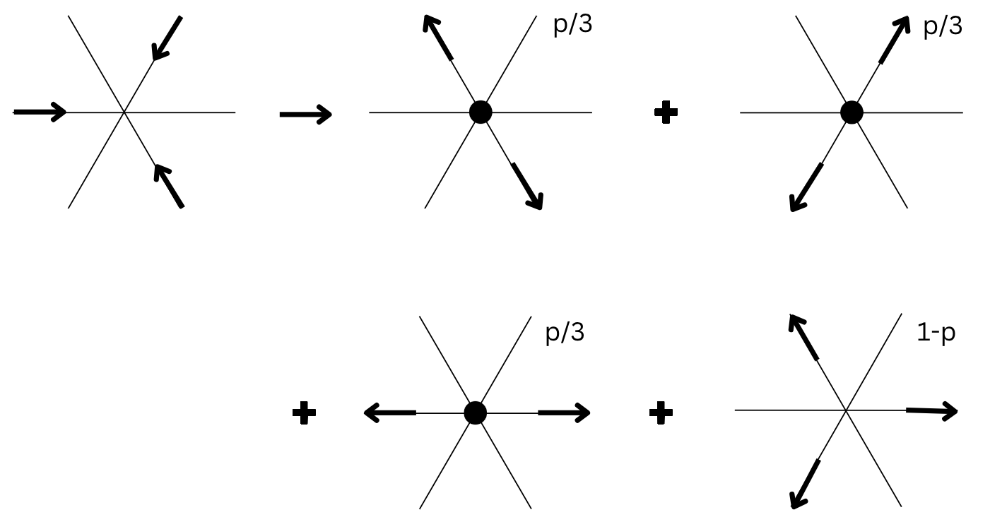}
  \caption{\label{fig:4} A three-particle collision event can lead to different outcomes depending on the value of $p$. Collisions conserve energy and momentum in the absence of energy dissipation ($p = 0$). However, when p $>$ 0, there are three possible outcomes, each with a probability of $p/3$. In each outcome, one of the particles experiences energy dissipation. Note that the hexagonal lattice configuration limits the direction of movement to the given array of possibilities.}
\end{figure}

In LGA models, particle interactions are typically assumed to be elastic collisions, which works well to model ideal gases but not granular materials~\cite{Kim2017}. Several research experiments have found that granular particle collision can be elastic or show energy dissipation
due to particle crushability or grain roughness that varies from one particle to another~\cite{ElShamy2014, 
Mukwiri2017, Zhao2008}.
To address this possibility, we adapt the collision model proposed by Herrmann et al.~\cite{Herrmann1996}
and introduce an energy dissipation parameter, $p$, ranging from $0$ to $1$---a value of $0$ corresponds to perfectly elastic collisions, while a value of $1$ signals fully inelastic collisions with complete energy dissipation. Since the roughness of particles is not known in advance, an additional parameter $p$ accounts for probabilistic energy dissipation in collisions, following the rules depicted in Figs.~\ref{fig:2}-\ref{fig:4}. Additionally, we implement mirror deflection to simulate the system's behavior when particles collide with walls.

\subsection{\label{subsec:friction-effects} The Onset of the Jamming Transition}

Experimental studies have consistently demonstrated the critical role of the formation of an arch structure at the narrow opening of a hopper in initiating jamming transitions~\cite{Kiwing2005}. This arch structure results from the friction between the particles and the hopper walls, creating a barrier that obstructs the material flow. 

\begin{figure}[t]
  \includegraphics[width=\columnwidth]{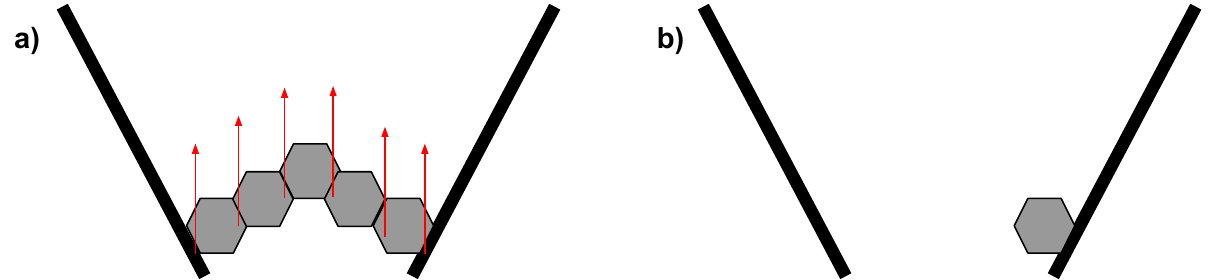}
  \caption{\label{fig:5} Illustration of an arch-like structure formation as a result of the upward frictional forces caused by the compression from both sides of the walls in 5a. In contrast, 5b illustrates a scenario where the particle does not experience friction from the wall due to the absence of compression from the left side.}
\end{figure}

To model friction within a lattice gas system, we have carefully designed a friction rule that governs the interaction between particles and their surrounding environment. 
Firstly, the walls exert a force in the opposite direction to the movement of particles. Secondly, particles can transmit frictional forces when they come into contact. However, for the friction force to take effect, particles must be compressed from both sides, creating a state of compression. Figure~\ref{fig:5}a shows particles arranging themselves into an arch-like structure, with arrows indicating the upward-pointing frictional forces due to the compression of particles toward the walls. The friction force is transmitted upwards through the compressed particles, ultimately leading to the structure acting as a barrier, impeding the flow of particles.

\section{\label{sec:level3} Simulation Results}

\subsection{\label{subsec:level3-1} Entropy of the LGA Model}

\begin{table*}[ht]
\caption{\label{tab:1}Parameter Values for Each Simulation}  

\centering 
\begin{tabular}{l c c c c} 
\hline 
& $g$ & $p$ & Grid H$\times$W & Density \\ [0.5ex] 
\hline 
Particles in a box (Figs. 6--7) & 0 & 0 & 100$\times$100 & 10\% \\ 
Density Waves (Figs. 8--9) & 0.2 & 0.1 & 200$\times$50 & 70\% \\
Narrow Pipe (Figs. 10--11) & 0.2 & 0.1 & 200$\times$10 & 85\% \\
Jamming (Figs. 12--13) & 0.2 & 0.1 & 200$\times$20 & 5\%--100\% \\ [1ex] 
\hline 
\end{tabular}
\label{table:nonlin} 
\end{table*}

\begin{figure}[t]
  \includegraphics[width=\columnwidth]{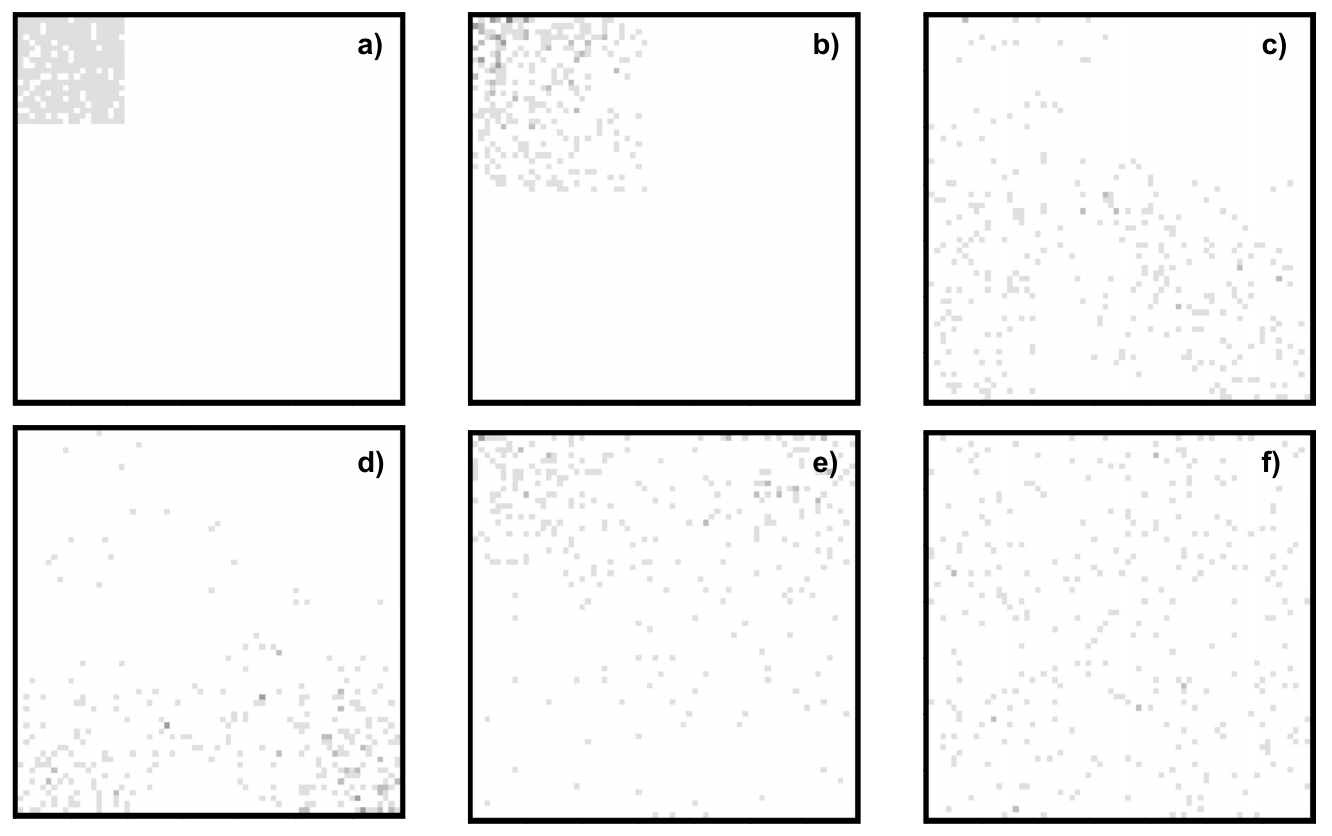}
  \caption{\label{fig:6} The time evolution of the particles after being confined to a small region in a box. A video of the simulation can be consulted \href{https://drive.google.com/file/d/1VPw-MfMx_v8i3npaYBgTmgFj4G6l_7kv/view}{here}.}
\end{figure}

In the absence of gravitational and frictional effects, the model is suitable for simulating standard fluid flows and is expected to adhere to the principles of thermodynamics. Notably, the second law of thermodynamics should apply, indicating that the system's entropy, when out of equilibrium, should increase with time. This implies that the number of possible microstates available to the system should progressively grow. We initiated the simulation with a low entropy state to verify this, as shown in Fig.~\ref{fig:6}a. The parameters used for simulation in this section and for the following sections are listed in Tab.~\ref{tab:1}.

The system, depicted in Figure~\ref{fig:6}, starts with a low entropy state, where all particles are placed in one corner of the box. 
As the system evolves, the particles will collide and occupy a larger portion of the available volume (Figs~\ref{fig:6} b-f). Consequently, the number of possible microstates at equilibrium will significantly exceed the restricted microstates confined to the $20 \times 20$ region, increasing entropy. To quantify the entropy of this system, we employ the Shannon entropy, defined as follows:
\begin{equation}
\label{eq:1}
H(x) = -\sum_{i=1}^{n} P(x_i) \log(P(x_i))
\end{equation}
In the context of information entropy, $x$ represents a random variable, while $x_i$ represents the possible outcomes associated with that random variable. The term  $P(x_i)$ denotes the probability of a specific outcome $x_i$. In the present study, the equation is applied by making $x$ akin to a random indicator variable: $1$, indicating the cell is occupied by a particle, or $0$, showing the cell is empty. To compute the probability $P(x_i)$, we conduct 50 distinct simulations, each spanning 500 steps. This generates an ensemble of 50 microstate configurations for each iteration. The probability $P(x_i)$ for a given cell $x_i$ in a particular iteration is then computed as the average occupancy frequency of that cell over the 50 distinct simulations. This average is determined by dividing the number of times cell $x_i$ is occupied by the particle by the total number of simulations (50). After computing the probabilities $P(x_i)$ for all cells, they are summed using equation (\ref{eq:1}). This procedure is repeated for each iteration in the model's evolution.

\begin{figure}[t]
  \includegraphics[width=\columnwidth]{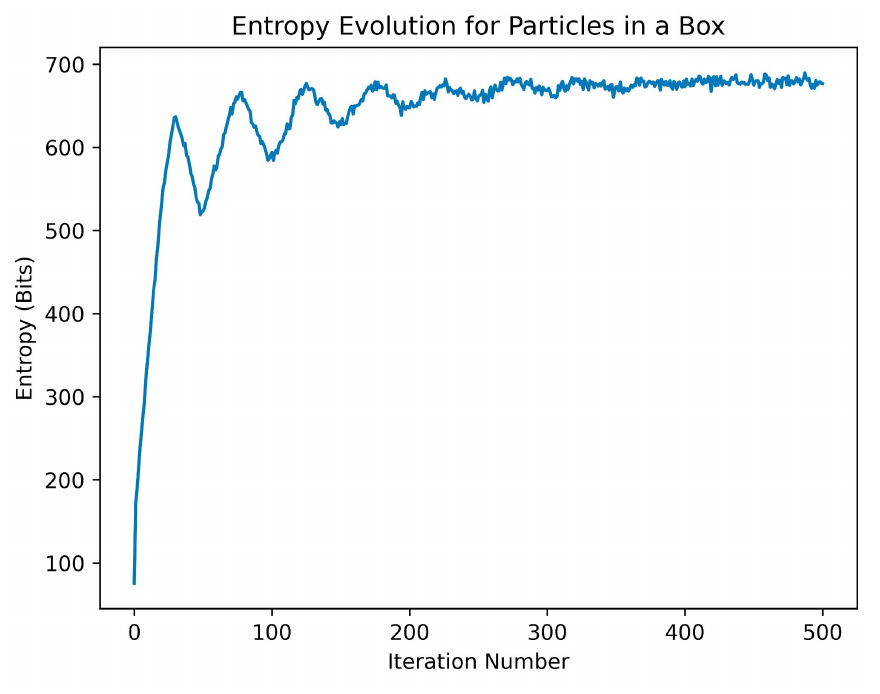}
  \caption{\label{fig:7}The entropy over time for the system initialized with a non-equilibrium state (Fig.~\ref{fig:6}). }
\end{figure}

In Fig.~\ref{fig:7}, we show the time evolution of entropy in the system that accompanies the stages depicted in Fig.~\ref{fig:6}. A transient decline in entropy is allowed---rapid entropy shifts can occur in out of equilibrium states. We further confirm the finding by Tribel and Boon that the monotonous increase in LGA entropy is not guaranteed~\mbox{\cite{Tribel1997}}. This fluctuating behavior persists equilibrium is reached, and a homogeneous distribution of particles across the entire box and a high entropy ensue.

\subsection{\label{subsec:level3-2} Density Waves in a hopper flow}

 Behringer et al. initially documented density waves in 1989 \cite{Fagert1989}. Employing a Plexiglas hopper with adjustable width, the team captured gravity-driven sand flow using X-ray imaging through digital fluoroscopy. They examined two sand types: rough (resulting in inelastic collisions due to crushability and shear strain) and smooth (with predominant elastic collisions). These experiments established insights into density wave formation and behavior in granular materials.

\begin{figure}[!ht]
\centering
  \includegraphics[height=0.5\textheight]{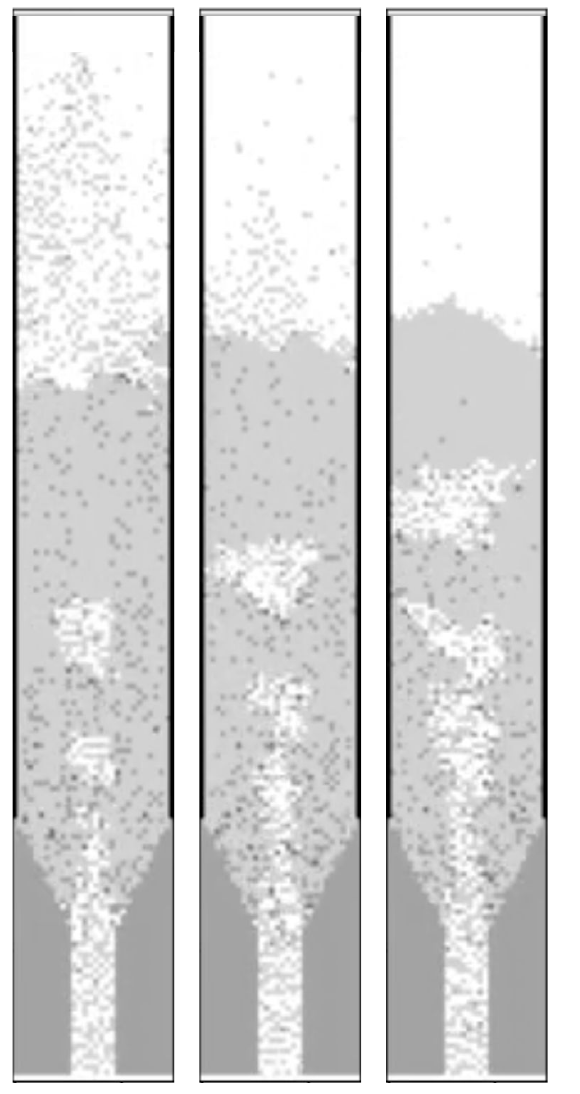}
  \caption{\label{fig:8} Snapshots of the sand flow in a hopper taken from the model simulation. The three consecutive snapshots, separated by a few iterations, illustrate the temporal evolution of the particle discharge process. A video of the simulation can be consulted \href{https://drive.google.com/file/d/1RIyhdxAj1mdDkLqn-jKdBHR9WnargcpH/view}{here}.}
\end{figure}

We successfully reproduced the phenomenon of density waves observed in the experimental study through simulations. As depicted in Fig.~\ref{fig:8}, the flow forms a central low-density region positioned directly above the outlet. This central region serves as the primary pathway for the flow, while the adjacent regions near the walls exhibit limited movement. We also observed the emergence of low-density regions propagating upwards between the central region and the walls, contrary to the downward sand flow. These propagating density waves persist and are followed by additional waves until they reach the system's top, eventually collapsing. This simulation-based observation aligns with the previous experimental findings on the propagation of density waves~\cite{Fagert1989}.

\begin{figure}[t]
  \includegraphics[width=\columnwidth]{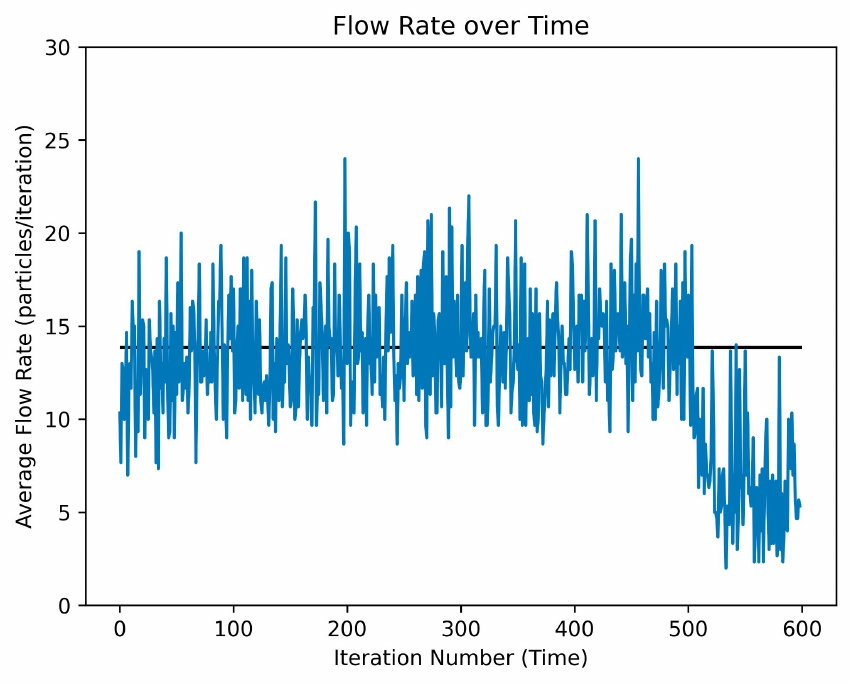}
  \caption{\label{fig:9}  Flow rate of the granular materials measured leaving the hopper outlet. The data is averaged over ten iterations, showing a roughly constant flow rate (indicated by the black horizontal line) until a decrease occurs towards the end.}
\end{figure}

The flow rate evolution at the hopper outlet, measured by counting the number of particles leaving the outlet every iteration, is depicted in Fig.~\ref{fig:9}. It is characterized by a consistent flow rate during the particle discharge, followed by a non-linear reduction towards the end that can be attributed to the decreasing particle density towards the end. 
The observed flow rate behavior aligns with experimental findings~\cite{Behringer1990, Baxter1991}.

\subsection{\label{subsec:level3-3} Density Waves in narrow pipes}

In 1994, Peng et al. provided a distinctive perspective on density wave propagation, relevant to industrial applications \cite{Peng1994}. Their study, utilizing narrow pipes instead of conventional hoppers, examined smooth and rough sand. Density wave propagation exclusively occurred with rough sand. The underlying reason is the inelastic collisions experienced by rough sand particles during descent. These collisions lead to energy dissipation, causing certain particles to lose velocity. Consequently, these decelerated particles lag behind, triggering successive collisions and more energy loss. This cascading effect resembles shock waves in traffic, where one car's abrupt stop triggers a chain reaction of deceleration among preceding vehicles.

\begin{figure}[!ht]
\centering
  \includegraphics[height=0.4\textheight]{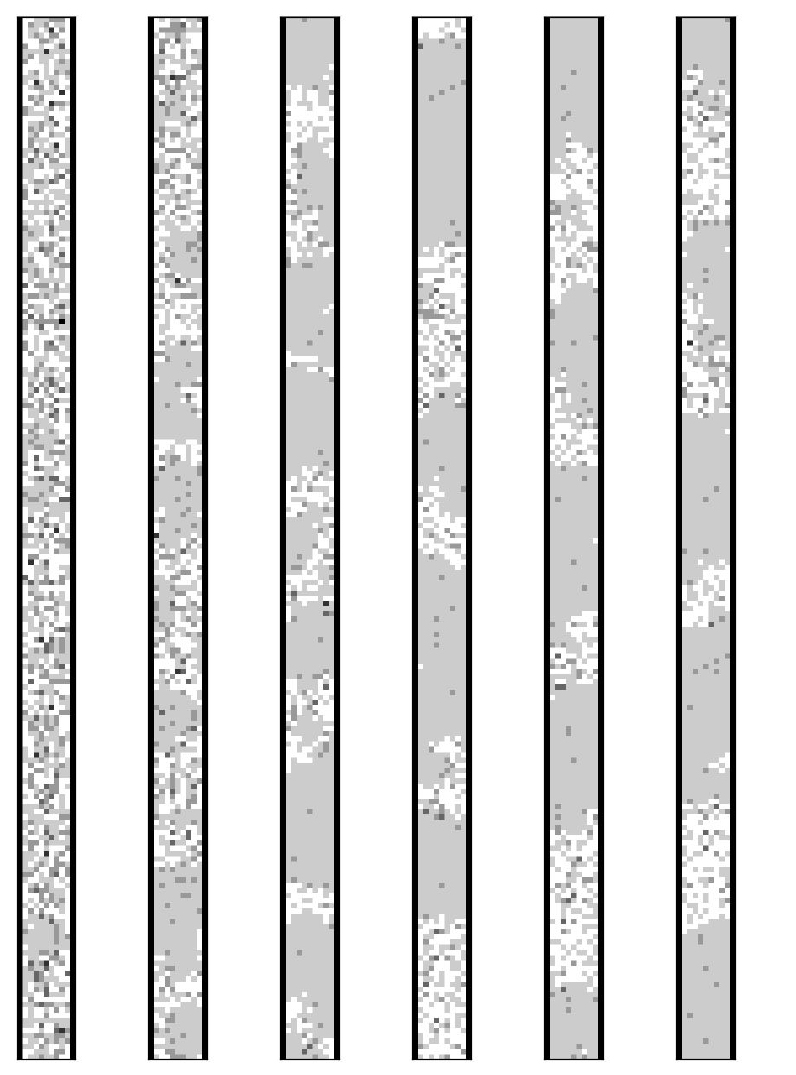}
  \caption{\label{fig:10} The set of figures, from left to right, shows the evolution of the flow in a narrow pipe and under periodic boundary conditions. Note the formation of shock waves. A link to the simulation video can be consulted \href{https://drive.google.com/file/d/1SLK07UM7Nnj-PoKoTeC3KrgavXdE2Sql/view}{here}.}
\end{figure}

Our model successfully replicated the results when we adjusted the simulation settings to a narrow pipe and replaced the hopper outlet with periodic boundary conditions. This adjustment allowed particles that flowed through the bottom of the pipe to re-enter the system through the top, ensuring a continuous flow. The simulation results are presented in Fig.~\ref{fig:10}.

\begin{figure}[t]
  \includegraphics[width=\columnwidth]{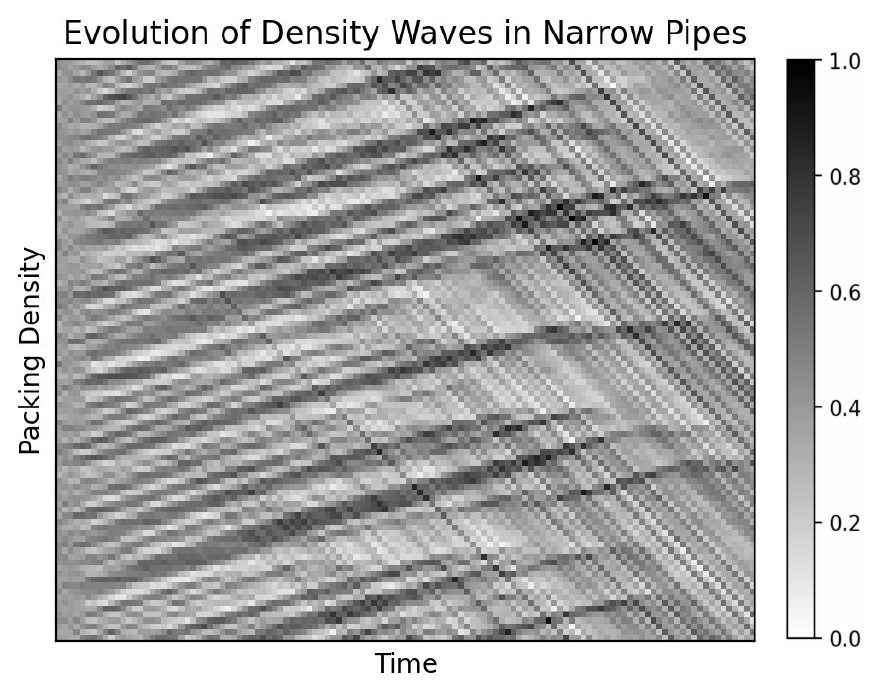}
  \caption{\label{fig:11} Sequential snapshots illustrating density wave propagation in a narrow pipe over time. Each snapshot represents a vertical segment of the pipe, with density depicted using grayscale colors (darker shades indicating higher densities). }
\end{figure}

Fig.~\ref{fig:10} illustrates a notable phenomenon---the emergence of a shock wave of high density and propagation in an upward direction. This behavior is prominently observed in narrow pipes, as the occurrence of inelastic collisions results in the halting of numerous particles from flowing, consequently leading to the formation of the density wave (or shock wave). Figure~\ref{fig:11} provides a comprehensive visualization of the temporal propagation of density waves, following a methodology similar to that employed by Peng et al.~\cite{Peng1994}. Multiple snapshots were taken from the narrow pipe at different time intervals to generate the figure. Each snapshot involved dividing the pipe into small vertical segments and calculating the density within each segment. The density values were then represented using a grayscale scheme, with darker regions indicating higher densities. These snapshots were sequentially arranged from left to right, creating the impression of multiple narrow pipes aligned horizontally, with time progressing from left to right. 

The graph serves as a clear demonstration of the formation and movement of density waves. The initial state, depicted on the left side of the graph, portrays a semi-uniform density distribution. As time elapses from left to right, density waves' emergence and upward propagation become increasingly evident. It is worth noting that once the density wave reaches the top of the graph, it reappears at the bottom due to the periodic boundary representation implemented in the simulation. After $10,000$ iterations near the right side of the figure, the number of density waves reduces as the voids merge into larger voids.

\subsection{\label{subsec:level3-4} Jamming Transition}

As mentioned above, the jamming transition occurs when an arch structure forms at the narrow opening of a hopper, obstructing the particle flow. This structure originates from the friction between the particles and the hopper walls, effectively blocking the granular flow. Multiple experimental studies, such as the one by Kiwing et al.~\cite{Kiwing2005} have explored the underlying mechanisms governing this process highlighting the role of the specific hopper structure. The emergence of this configuration is stochastic in nature and dependent on particle's phase space.

Furthermore, these investigations shed light on the role of density in jamming transition. Higher densities are associated with increased jamming transition, as they entail more particles and a more extensive set of possible microstates, which 
significantly enhances the probability that one of the configurations will lead to the formation of an arch and subsequent jamming. Consequently, the jamming transition is not solely contingent upon reaching a critical density value but becomes more likely above this critical density threshold. This distinction sets the jamming transition apart from traditional phase transitions, such as the water-ice transition, typically occurring at specific critical values of control parameters like temperature or pressure. Our simulations identified a critical density threshold above which jamming transition is more likely to occur, as illustrated in Fig.~\ref{fig:13}.

\begin{figure}[t]
  \includegraphics[width=\columnwidth]{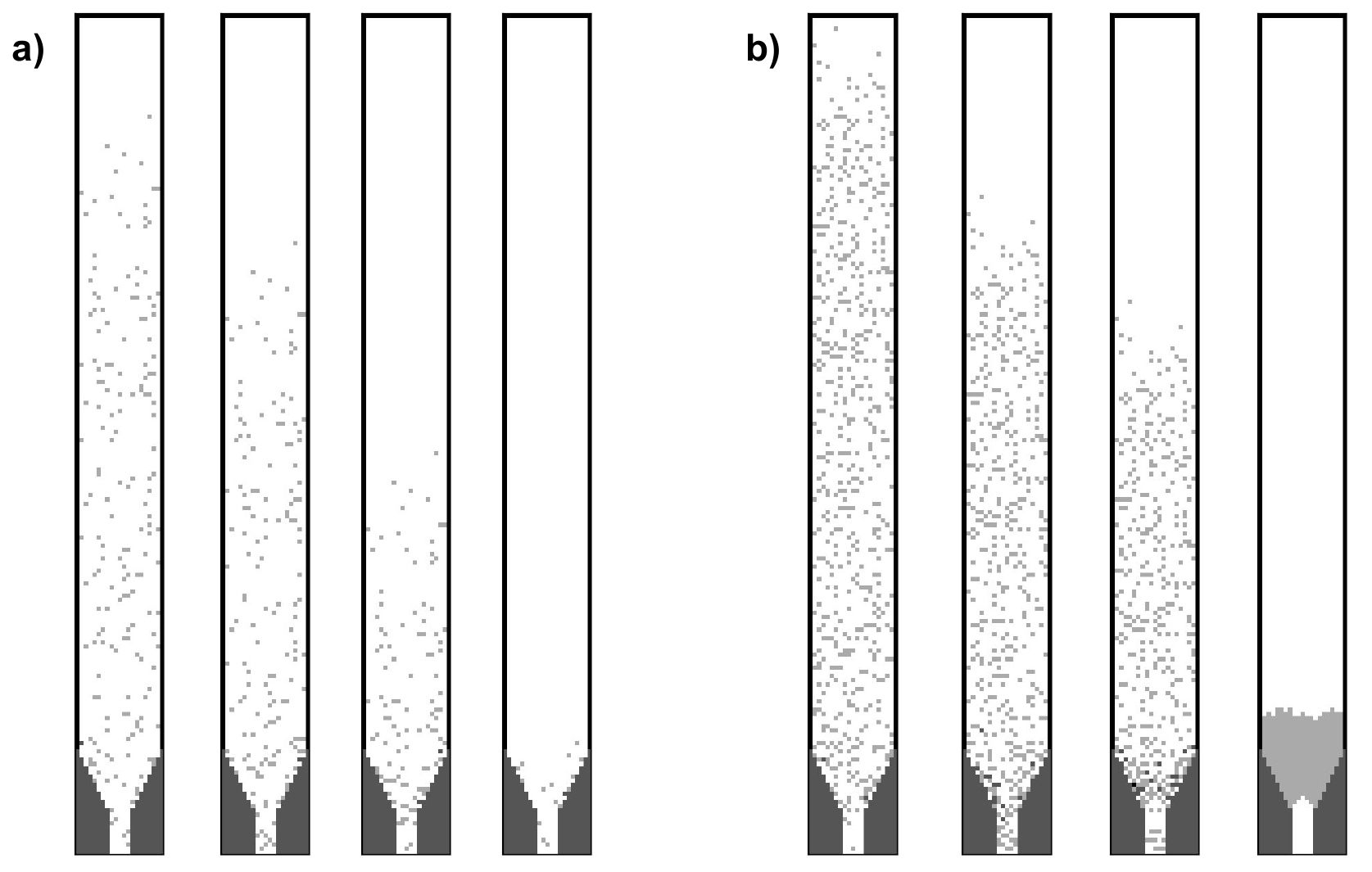}
  \caption{\label{fig:12}  Flow evolution in a narrow-outlet hopper. Fig. 12a is initialized with $15\%$ packing density for which no jamming occurs, while Fig. 12b is initialized with $30\%$ packing density for which jamming occurs. A video of the simulation can be consulted \href{https://drive.google.com/file/d/1a3qt3xV2IbL_RJrSNKHu4-HF_ZRZhD0M/view}{here}.
 }
\end{figure}

The simulation depicted in Fig.~\ref{fig:12} involved initializing a narrow hopper with varying packing densities of particles, followed by measuring the resulting flow rate. To achieve this, a predetermined number of particles corresponding to each density was randomly distributed within the hopper. The simulation was then executed for 5000 iterations. At each density, the flow rate was quantified as the average number of particles exiting through the hopper outlet per unit of time. To ensure the robustness of the results, this process was repeated $30$ times for each density. The reported flow rate in Fig.~\ref{fig:13} is the average flow rate from the $30$ simulations for each density value, with error bars indicating the 95\% confidence intervals.

As expected, Fig.~\ref {fig:13} demonstrates a persistent linear relationship between the particle density and the flow rate until reaching a density of approximately $20$\%, beyond which the system contains enough particles to increase the probability of arch formation, reducing the flow rate. As the density approaches $100$\%, the box becomes densely packed with particles, resulting in a near-certain arch formation and a complete halt in particle flow. It should be noted that, statistically, the flow rate will not reach zero as a few particles may still exit the system before the arch forms.

\begin{figure}[t]
  \includegraphics[width=\columnwidth]{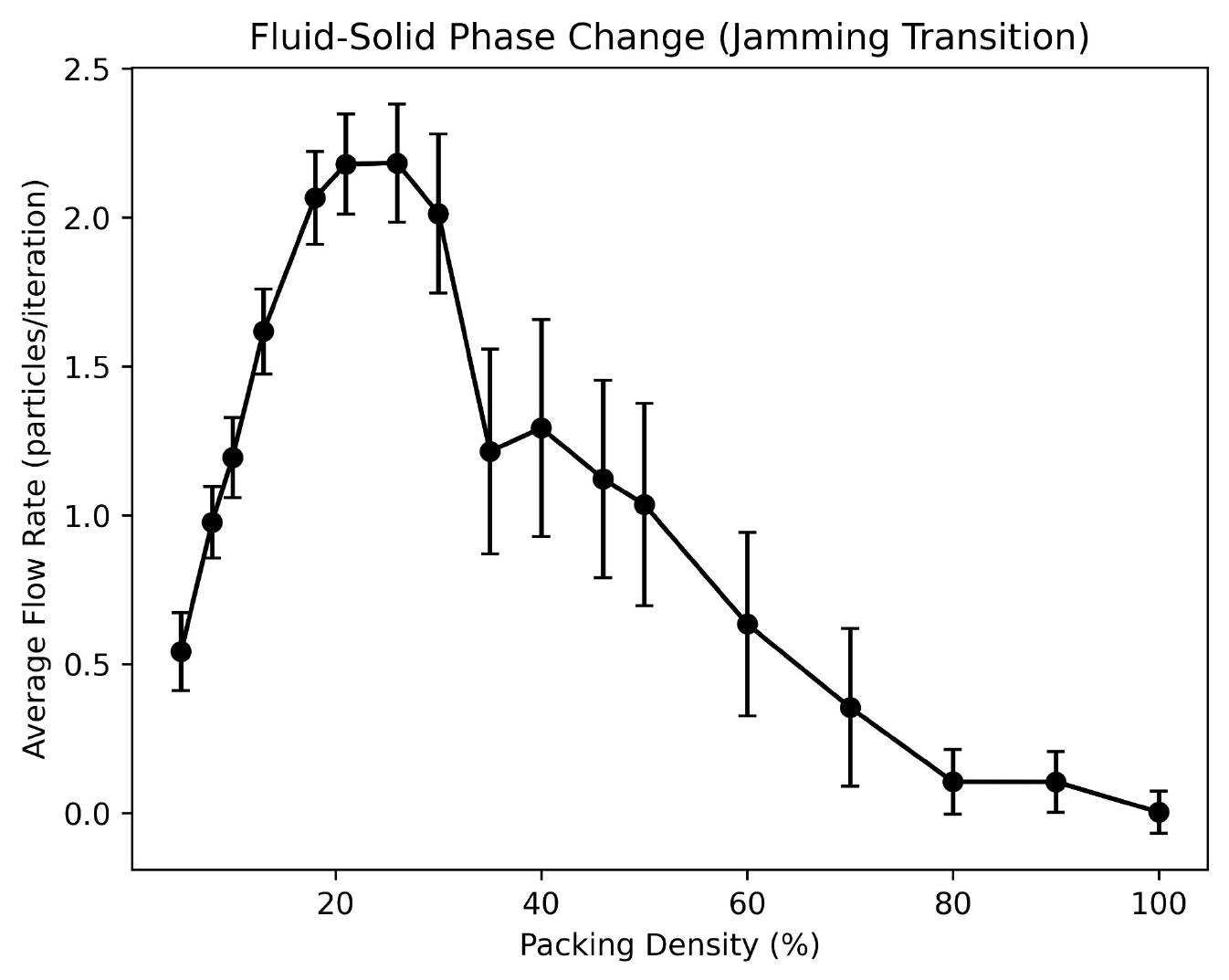}
  \caption{\label{fig:13} The graph illustrates the flow rate measured at different densities and the onset of the jamming transition. The model was run for 30 different simulations at each density, each running for 5000 iterations. The average flow rate was reported with 95\% confidence intervals.}
\end{figure}

We also explored jamming and flowing outcomes for multiple particle densities. Jamming is relatively rare for densities below $20$\%, with infrequent flowing events for densities above $20$\%. This probabilistic nature of jamming indicates that higher densities increase the likelihood of its occurrence, again corroborating previous experimental findings~\cite{Kiwing2005}.

\section{\label{sec:level4} Conclusion}

This study reports on the first successful simulation of granular materials using an LGA framework. 
The model accurately replicates critical empirical observations by incorporating gravitational effects, energy dissipation in particle collisions, and frictional interactions. 

The main novelty of this research lies in the reproduction of the jamming transition, previously only observed 
experimentally. We recognized the probabilistic nature of the critical density threshold for arch formation 
demonstrating how it is more likely at higher densities. 
Our tailored LGA model also succeeds in reproducing the flow rate evolution and density wave formation at the hopper outlet. The simulated non-linear reduction in flow rate towards the end of the discharge process suggests the role of density fluctuations in arch formations.

These findings further our understanding of granular dynamics, providing valuable insights into the complex behavior of granular materials. Based on these results, one natural direction to follow up would be to delve into how the jamming transition in granular materials might affect the propagation and behavior of density waves. Investigating the interplay between the jamming transition and the emergence of density waves could unveil intriguing connections between these two phenomena, potentially shedding light on the underlying mechanisms that govern both processes. This venue of research could contribute to a more comprehensive understanding of the complex behaviors exhibited by granular materials near the jamming transition and their implications for various practical applications.\\

\noindent{\textbf{ORCID}}\\
\newcommand{\OrcidLink}[1]{\href{https://orcid.org/#1}{#1}}
\noindent Mohamed Gaber:        \OrcidLink{0009-0004-9562-1244}\\
\noindent Raquel H. Ribeiro:    \OrcidLink{0000-0002-5434-252X}\\
\noindent Janek Kozicki:        \OrcidLink{0000-0002-8427-7263}


\bibliography{manuscript.bib}

\end{document}